\documentclass{article}
\usepackage{amsmath}
\usepackage{graphicx}
\usepackage{xcolor}
\DeclareRobustCommand{\xapm}{a^p \! \! \! \mod m}
\usepackage{authblk}
\begin{document}
\emergencystretch 2em
\setcounter{Maxaffil}{2}
\author{Frederick James}
\author{Lorenzo Moneta}
\affil{CERN, Geneva, Switzerland}
\date{3 March 2019}
\title{Review of High-Quality Random Number Generators  }

\maketitle

\begin{abstract}
This is a review of pseudorandom number generators (RNG's) of the 
highest quality, suitable for use in the most demanding Monte Carlo  
calculations.
All the RNG's we recommend here are based on 
the Kolmogorov-Anosov theory of mixing in classical 
mechanical systems, which guarantees under certain conditions 
and in certain asymptotic limits, that points on 
the trajectories of these systems can be used to produce
random number sequences of exceptional quality.
We outline this theory of mixing and establish criteria for 
deciding which RNG's are sufficiently good approximations
to the ideal mathematical systems that guarantee highest quality.
The well-known RANLUX (at highest luxury level) 
and its recent variant RANLUX++ are seen to
meet our criteria, and some of the proposed versions of MIXMAX can be
modified easily to meet the same criteria.

\end{abstract}

\section{High-quality random numbers}
We are concerned here with  pseudorandom number generators (RNG's),
in particular those of the highest quality.
It turns out to be difficult to find an operational definition of randomness
that can be used to measure the quality of a RNG, that is the degree of 
independence of the numbers in a given sequence, or to prove that they 
are indeed independent.
The situation for traditional RNG's 
(not based on Kolmogorov-Anasov mixing)
is well described by Donald Knuth in \cite{knuth}.
The book contains a wealth of information about random number 
generation, but nothing about where the randomness comes from,
or how to measure the quality (randomness) of a generator.
Now with hindsight, it is not surprising that all the 
widely-used generators described there were later found 
to have defects (failing  tests of randomness and/or
giving incorrect results in Monte Carlo (MC) calculations), 
with the notable exception of RANLUX, which
Knuth does mention briefly in the third edition, but without describing
the new theoretical basis.

\subsection{The need for high quality.}
   High-level scientific research, like many other domains, has become 
dependent on Monte Carlo calculations, both in theory and in all
phases of experiments.  It is well-known that the MC method is used 
primarily for calculations that are too difficult or even impossible to
perform analytically, so that our science has become dependent
to a large extent on the random numbers used in extensive
MC calculations.  But how do we know if those numbers are 
random enough?  In the early days (1960's) the RNG's were so poor
that, even with the very slow computers of that time, their defects 
were sometimes obvious, and users would have to find a better 
RNG before continuing.  That usually meant using another linear 
congruential generator (LCG) with a different multiplier.  
When the result looked good, it was assumed to be correct, 
and often it probably was correct, but we will never know.  

As computers got faster and RNG's got longer periods, the situation 
evolved quantitatively, but still unacceptable results were occasionally
obtained and of course were not published.
Until  1992, when the famous paper of 
Ferrenberg, Landau and Wong \cite{ferrenberg}
showed that the RNG considered at that time to be the best 
was giving the wrong answer to a problem in phase transitions,
while the older RNG's known to be defective gave the right answer.
Since most often we don't have any independent way to know the 
right answer, it became clear that empirical testing of RNG's, 
at that time the only known way to verify their quality, was not
good enough.  Fortunately, the particular problem which was
detected by Ferrenberg et. al. was soon solved by Martin L\"uscher
 (in \cite{luscher}), but it became clear that if we were to have 
 confidence in MC calculations, we would need a better way to 
ensure their quality.
Fortunately, the theory of K-systems offers this possibility.

The experience gained from developing, using and discovering defects
in many RNG's has taught us some lessons which we summarise here
(they are explained in detail in \cite{knuth}):
\begin{enumerate}
\item{The period should be much longer than any sequence 
that will be used in any one calculation, 
but a long period is not sufficient to ensure lack of defects.}
\item{Empirical testing can demonstrate that a RNG has defects 
(if it fails a test), but passing any number of empirical tests can never 
prove the absence of defects.}
\item{Making an algorithm more complicated (in particular, 
combining two or more methods in the same algorithm) may 
make a better RNG, but it can also make one much worse than
a simpler component method alone if the component methods
are not statistically independent.}
\item{It is better to use a RNG which has been studied, 
whose defects are known and understood, than one which looks
good but whose defects are not understood.}
\item{There is no general method to determine how good a RNG
must be for a particular MC application.   
The best way to ensure that a RNG is good enough for a given 
application, is to use one designed to be good enough for 
all applications.}
\end{enumerate}

\section {The Theory of Mixing in Classical Mechanical Systems}

It has been known, at least since the time of Poincar\'e, that classical
dynamical systems of sufficient complexity can exhibit chaotic behaviour,
and numerous attempts have been made to make use of this
``dynamical chaos'' to produce random numbers by numerical 
algorithms which simulate mechanical systems.  
It turns out to be very difficult to find an approach which produces a
practical RNG, fast enough and accurate enough for general MC
applications. 
To our knowledge, only two such attempts have been successful,
both based on the same representation and theory of dynamical 
systems. 

\paragraph{}
This theory grew out of the study of the asymptotic behaviour of 
classical mechanical systems that have no analytic solutions, 
developed largely in the
 Soviet Union around the middle of the twentieth century by 
 Kolmogorov, Rokhlin, Anosov, Arnold, Sinai and others. 
 See, for example, Arnold and Avez \cite{arnold}  for the theory.
 At the time,  these mathematicians  were certainly not thinking 
 of RNG's, but it turns out that their results can be used 
 to produce sequences of random numbers that have some of the
 properties of the trajectories of the dynamical systems.
 (See  Savvidy 1991 \cite {sav91} and Savvidy 2016 \cite{george2016}.)
The property of interest here is called Mixing, and is usually associated 
with the names Kolmogorov and Anosov. 
Mixing is a well-defined concept in the theory, 
and will be seen to correspond quite exactly
to what is usually called independence or randomness.

\paragraph{}
The representation of dynamical systems appropriate for our purposes is
the following:
            \begin{equation}  \label{eq:eq1} 
       a(i+1) =  A \times a(i)  \mod 1 
            \end{equation}
where $a(i) $ is the $N$-vector of real numbers 
specifying completely the state of the system
at time $i$, and $A$ is a (constant) $N \times N$ matrix which can be 
thought of as representing the numerical solution to the equations of motion.
The $N$-dimensional state space is a unit hypercube which because of
the \emph{modulo} function becomes topologically equivalent to an 
$N$-dimensional torus by identifying opposite faces. 
The vectors $a$ represent points along the continuous 
trajectory of the abstract dynamical system in N-dimensional
phase space.

All the elements of the matrix $A$  are integers, 
and the determinant of $A$ must be one.  
This ensures that $A$ is invertible and 
the elements of $A^{-1}$ are also integers. 
The theory is intended for high- (but finite) dimensional systems 
($ 1 \ll N < \infty$). In practice $N$ will be between 8 and a few hundred.

\subsection{Mixing and the ergodic hierarchy}
Let $a(i)$ and $a(j)$ represent the state of the
dynamical system at two times $i$ and $j$.  
Furthermore, let $v_1$ and $v_2$ be
any two subspaces of the entire allowed space of points $a$, with
measures (volumes relative to the total allowed volume) respectively
$\mu(v_1)$ and $\mu(v_2)$.  Then the dynamical system is said to be 
a 1-system (with 1-mixing) if
 $$ P(a(i)  \in v_1 ) = \mu(v_1) $$ 
and a 2-system (with 2-mixing) if
 $$ P(a(i)  \in v_1 \; {\mathrm and} \; a(j) \in v_2 ) = \mu(v_1) \mu(v_2) $$ 
for all $i$ and $j$ sufficiently far apart, and for all subspaces $v_i$.
Similarly, an $n$-system can be defined for all positive integer values $n$.
We define a  zero-system as having the \emph {ergodic} property 
(coverage),  namely
that the state of the system will asymptotically come arbitrarily close 
to any point in the state space.  

\paragraph{}
Putting all this together, we have that asymptotically:
\begin{itemize}
\item A system with zero-mixing covers the entire state space.
\item A system with one-mixing covers uniformly.
\item A system with two-mixing has $2 \times 2$ independence of points.
\item A system with three-mixing has $3 \times 3$ independence.
\item etc.
\end{itemize}
Finally, a system with $n$-mixing for arbitrarily large values of $n$ is 
said to have K-mixing and is a K-system.
It is a result of the theory that the degrees of mixing form a 
\emph{hierarchy} \cite{berkovitz}, 
that is, a system which has $n$-mixing 
for any value of $n$ also has $i$-mixing for all $i<n$.
There are additional systems,
 in particular Anasov C-systems and Bernoulli B-systems
which are also K-systems, but K-systems are
sufficient for our purposes.

Now the theory tells us that a dynamical system represented by
equation (\ref{eq:eq1}) will be a K-system if the matrix $A$ has 
determinant equal to one and eigenvalues $\lambda$, 
all of which have moduli $|\lambda_i | \neq 1$.

\subsection{The eigenvalues of A}
We have seen that to obtain K-mixing, none of the eigenvalues of $A$
should lie on the unit circle.  In fact, in order to obtain sufficient mixing 
as early as possible (recall that complete mixing is only an asymptotic property)
it is desirable to have the eigenvalues as far as possible from the 
unit circle.  

An important measure of this distance is the Kolmogorov entropy $h$:
$$   h =  \sum _{k:|\lambda_k | > 1} \ln|\lambda_k|  $$
where the sum is taken over all eigenvalues greater than 1.
[It is also equal to the sum over all eigenvalues less than 1, 
but then it changes sign.]  As its name implies, it is analogous to 
thermodynamic entropy as it measures the disorder in the system,
and it must be positive for an asymptotically chaotic system. 
[This actually follows from the definition if all $|\lambda| \ne 1$.]

Another important measure is the Lyapunov exponent, 
defined in the following section.

\subsection{The divergence of nearby trajectories} 
The mechanism by which mixing occurs in K-systems can be observed
``experimentally" by noting the behaviour of two trajectories which start 
at nearby points in state space. 
Using the same matrix $A$, let us start the generator from 
two different nearby points $a(0)$ and $b(0)$, separated by 
a very small distance $\delta(a(0),b(0))$. The distance $\delta$ is defined by:
\begin{equation} \label{eq:dist}
\begin{aligned}
   \delta(a,b) &= \max_{\kappa}d_\kappa,    \\  
  d_\kappa&=\min\bigl\{|a_\kappa-b_\kappa| \, , \, 1-|a_\kappa-b_\kappa|\bigr\},
\end{aligned}
\end{equation}
where $\kappa$ runs over the $N$ components of the vector indicated.
Note that the distance defined in this way is a proper distance measure
and has the property $0\leq \delta \leq 1/2$.
Now we use equation (\ref{eq:eq1}) on $a(0)$ and $b(0)$ to produce 
$a(1)$ and $b(1)$, and calculate  $\delta(a(1),b(1))$. 
Then we continue this process to produce two series of points $a(i)$
and $b(i)$, and a set of distances $\delta(i)$, for  $i = 1,2,3 \ldots $ until 
the $\delta$ reach a plateau at their equilibrium value 
which according to L\"uscher is $\delta = 12/25$ for RANLUX.  
Then it is a well-known result of the theory that,
 if A represents a K-system, the distances $\delta_i$ will diverge  
 exponentially with $i$, so that if plotted 
on a logarithmic scale, the points $\delta_i$ vs. $i$ should lie
on a straight line.  The inevitable scatter of points should be reduced by
averaging the $\delta$ over different starting pairs $(a(0), b(0))$ 
(the $b(0)$ must of course always be the same very small distance
from the $a(0)$, but in different directions).

The rate of divergence of nearby trajectories  ($\nu$, the
Lyapunov exponent)  is equal to the logarithm
of the modulus of the largest eigenvalue:
$$ \nu  =   \ln |\lambda|_{max} $$
which, for a K-system, should be the slope of the straight line 
described above.

\subsection{Decimation}
The asymptotic independence of $a$ and $b$ is guaranteed by 
the mixing (if it is a K-system), but as long as  $\delta(i)$ remains small,
$a(i)$ and $b(i)$ are clearly correlated.  
The point where  the straight line of divergence of nearby trajectories 
reaches the plateau of constant $\delta  $ indicates the 
number of iterations $m$ required to make the K-system 
``sufficiently asymptotic", in the  sense 
that the points $a$ and $b$ generated on the following iteration
are on average as far apart as independent points would be.
We may call this criterion the ``2-mixing criterion", since it
apparently assures that 2x2 correlations due to nearby
trajectories will be negligible.
The question whether this criterion is sufficient to eliminate
higher order correlations, is important 
and will be discussed below in connection with RANLUX++.

Some plots of divergence for real RNG's are given below.
If the K-system is used to generate random numbers, 
in order to eliminate the correlations due to nearby trajectories, 
after one vector $a_i$ is
delivered to the user, the following $m$ vectors 
$a_{i+1},a_{i+2}, \ldots a_{i+m}$ must be discarded before the next
vector $a_{i+m+1}$ is delivered to the user (decimation).  
It has become conventional to characterise the degree of decimation
by the integer $p$, defined such that after delivering a vector of $N$
random numbers to the user, $p-N$ numbers are skipped before the next 
$N$ numbers are delivered \cite{luscher,sibidanov}. 

\section{From the Theory to the Discrete RNG}
Equation (\ref{eq:eq1}) will be used directly to generate random numbers,
where $a(0)$ will be the $N$-dimensional seed, and each successive 
vector $a(i) $ will produce $N$ random numbers.
However, in the theory, $a$ is a vector of real numbers, 
continuous along the unit line. 
The computer implementation must approximate the real line by 
discrete rational numbers, which is valid 
provided the finite period is sufficiently long,
and the rational numbers are sufficiently dense,
so that the effects of the discreteness are not detectable.  
Thus the computer implementation has access only to 
a rational sublattice of the continuous state space, and
we must confirm that the discrete approximation
preserves the mixing properties of the continuous K-system.  
Fortunately, the divergence of nearby trajectories offers this possibility, 
since a theorem usually attributed to Pesin \cite{berkovitz} states 
that a dynamical system 
has positive Kolmogorov entropy and is therefore K-mixing 
if and only if nearby trajectories diverge exponentially. 
Then we can expect the discrete system to be K-mixing only if
the same condition is satisfied.

\subsection {The Period}
The most obvious difference between continuous and discrete systems
is that continuous systems can have infinitely long trajectories,
whereas a computer RNG must always have a finite period.
This means that a trajectory `eats up' state space as it proceeds, 
since it can never return to a state it has previously occupied without
terminating its period.
The fact that the available state space becomes progressively smaller
indicates necessarily a defect in the RNG, but this defect is easily seen 
to be undetectable if the period is long enough.
According to Maclaren \cite{NMM}, when the period is $P$, 
using more than $P^{2/3}$ numbers from the sequence 
``leads to excessive uniformity compared to a true random sequence."
This is compatible with the RNG folklore which sets the usable limit
at $\sqrt{P}$. 

\subsection{Summary of Criteria for Highest Quality.}
We will consider as candidates for highest quality RNG's 
only those based on the theory of chaos in classical mechanical 
systems, for the simple reason that we know of no other class 
of systems which can offer -- even in some limit which may 
or may not be attainable --  the uniform distribution 
and lack of correlations guaranteed by k-mixing.  
To be precise, our criteria are the following:
\begin{enumerate}
\item 
         {Matrix A must have eigenvalues away from the unit circle, and determinant =1.}
 \item        
         {Kolmogorov entropy must be positive (follows from the above).}
\item 
         { Discrete algorithm must have points sufficiently dense to
         accurately represent the continuous system.}         
\item        
         { Divergence of nearby trajectories must be exponential
                    (follows from the above) .}
\item        
         { Decimation must be sufficient to assure that the average 
         distance between successive vectors is the expected distance
          between independent points.} 
\item        
          {Period must be long enough, $>10^{100} $ }
\item        
      { Some practical criteria: double precision available, portable, repeatable, independent  sequences possible.}

\end{enumerate}

\section{The High-quality RNG's: 1. RANLUX}

The first widely-used RNG to offer reliably random numbers was
Martin L\"uscher's RANLUX, published in 1994.  
He considered the RNG
proposed by Marsaglia and Zaman \cite{MZ}, installed many years ago
at CERN with the name RCARRY, and now
known variously as RCARRY, SWB (subtract with borrow) or AWC
(add with carry).  
He discovered that, if the carry bit was neglected (see below),
this RNG had a structure that could be represented by equation
(\ref{eq:eq1}), and was therefore possibly related to a K-system. 

\subsection{The SWB (RCARRY) algorithm}
SWB operates on an internal vector of 24 numbers, 
each with a 24-bit mantissa, 
and each call to the generator produces one random number which
is produced by a single arithmetic operation (addition or subtraction)
operating on two of the numbers in the internal vector.  
Then this random number replaces one of the entries in the internal vector.
L\"uscher realised that if SWB is called 24 times in succession
it generates a 24-vector which is related to the starting 24-vector by 
(\ref{eq:eq1}), and he could determine the matrix $A$ which would
reproduce almost the same sequence as SWB.
The only difference would be due to the  ``carry bit"  which 
is necessary for attaining a long period, but affects only the least 
significant bit, so it does not alter significantly
the mixing properties of the generator. 
The carry bit is described in detail, both in the paper of 
Marsaglia and Zaman \cite{MZ}
and that of  L\"uscher \cite{luscher}.

\subsection{SWB: The Lyapunov exponent and decimation}
The eigenvalues of this ``equivalent matrix" were seen to
satisfy the conditions for a K-system, but the RNG nevertheless
failed several standard tests of randomness.
Plotting the evolution of the separation of nearby trajectories
immediately indicated the reason for the failure and how to fix it.
The Lyapunov exponent was bigger than one, as required, 
but not much bigger, so it would require considerable decimation
in order to attain full mixing.

Note that the reason for needing decimation has nothing to do with
the carry bit, or even with the discrete approximation to the 
continuous system.  It would in general be needed even for 
an ideal K-system with continuous  $a_i$,  simply because 
mixing is only an \emph{asymptotic} property of K-systems.  

\begin{figure}[t]

\includegraphics[width=\linewidth]{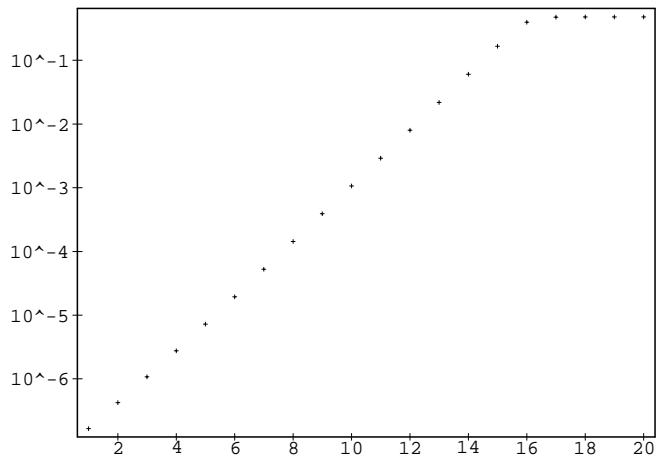}


\caption{The divergence of nearby trajectories for RANLUX}
\label{fig:1}
\end{figure}

The situation is shown in figure 1.  It can be seen first of all that
the first 15 points lie precisely on a straight line, indicating 
exponential divergence and therefore RANLUX has indeed
the mixing  properties of a K-system.  However we also see that,
after generating 24 numbers, it is necessary to throw away (decimate)
about $16 \times 24$ numbers before delivering the next 
24 numbers to the user, in order that the average separation between
nearby trajectories equals the expected separation between
independent points in phase space (the 2-mixing criterion).
Fortunately, throwing away numbers is typically about twice as fast as
delivering them, so the generation time is increased by only about
a factor 7, which is still negligible for many applications, since the
basic algorithm is very fast.  However, for those users who cannot afford
the extra time, RANLUX offers different degrees of decimation, 
known as ``luxury levels".  The lowest luxury levels provide very little or 
no decimation and are very fast; higher levels are slower but more 
reliable: and the highest level offers  sufficient decimation to eliminate
the obvious correlation due to nearby trajectories remaining too close.
Whether this decimation is sufficient to guarantee the disappearance of 
all higher order correlations will be discussed below.

\subsection{RANLUX: The spectral test.}

The spectral test is not exactly an empirical test, since it cannot be applied
to any sequence of numbers, but requires some knowledge of the 
RNG algorithm  \cite{knuth}.
We could therefore call it a semi-empirical test which can be used
on any linear congruential generator (LCG) with a known multiplier.
RNG's in this class have the well-known property that if $d$-tuples of
random numbers are used as coordinates of points in a $d$-dimensional 
hypercube, all the points lie on parallel hyperplanes which are sometimes  
spaced much more widely than would be the case for independent points.
Given the multiplier of the LCG and a value of $d$, the spectral test 
provides a figure of merit for the spacing of the hyperplanes.
This test is a full-period test and is generally considered to be
more powerful than the usual empirical tests.

   When the first papers on RANLUX were published, Tezuka,
L'Ecuyer and Couture \cite{tezuka} had already discovered that the
 algorithm  of Marsaglia and Zaman \cite{MZ} 
 (which was the basis for RANLUX) was in fact equivalent 
 to a linear congruential generator (LCG) with an enormous
multiplier.   
This fact was used by L\"uscher \cite{luscher} in order to apply 
the spectral test to RANLUX.
He shows that RANLUX passes the test at high luxury levels, 
but fails at the lower levels as expected.

\subsection{RANLUX: The period and the implementation}
So RANLUX is simply SWB with decimation, where the amount of
decimation is a parameter to be chosen by the user: the luxury level.  
The definitions of the luxury levels may depend on the implementation, 
so the user should consult local documentation.

The period of SWB is known exactly and is 
$\approx 5 \times 10^{171}$.  
This is easily seen to be many orders of magnitude more than 
all the world's computers could generate in the expected
lifetime of the earth, so any defects arising from the finiteness
of the state space should remain undetectable.  

The original implementation was in Fortran \cite{james}
 but RANLUX is now used 
mainly in the implementations in C, available in single precision for
24-bit mantissas, and in double precision for 53-bit mantissas, 
of which 48 are implemented and the last 5 are zeroes.  
These are available from L\"uscher's website (with documentation)
and are also installed in several program libraries including CERN
and the Gnu Scientific Library.
It is part of the C++ standard library.
There is also a version \cite{HJ} using integer arithmetic internally,
which may be faster on some platforms.
Because of the luxury levels, RANLUX is especially suited for testing
packages that test RNG's for randomness \cite{SB}.
If RANLUX passes the test at low luxury levels, then the test is
not very sensitive; if it fails at the highest luxury level, 
then it is likely that the test itself is defective.
\footnote{Several defects have been reported in RANLUX,
but in each case it turned out that the test procedure was incorrect 
or incorrectly applied.}

\section{The High-quality RNG's: 2. MIXMAX}
A few years before the publication of RANLUX, 
George Savvidy and his colleagues in Erevan \cite{sav91} were 
working on a different approach 
to the same problem using the same theory of mixing.
Their approach was to look for a family of matrices $A$ which could be 
defined for any dimension $N$, having eigenvalues far
from the unit circle and therefore large 
Lyapunov exponents for different values of $N$, in order to
reduce or even eliminate the need for decimation.

\subsection{MIXMAX: The matrix, the algorithm and decimation}
The family of matrices which they found was
 (for $N \ge 3$):

   \begin{equation}  \label{eq:savmat} 
  A = 
 \begin{pmatrix}
  1 & 1     & 1 & 1 & \cdots & 1 & 1 \\
  1 & 2     & 1 & 1 & \cdots & 1 & 1 \\
  1 & 3+s & 2 & 1 & \cdots & 1 & 1 \\
  1 & 4     & 3 & 2 & \cdots & 1 & 1 \\
  \vdots  & \vdots  & \vdots & \vdots & \ddots  &\vdots &  \vdots &\\
  1 & N-1 & N-2 & N-3 & \cdots & 2 & 1 \\
  1 & N & N-1 & N-2 & \cdots & 3 & 2 \\
 \end{pmatrix}   
  \end{equation}
  
\noindent  
where the ``magic" integer $s$ is normally zero, but for some values 
of $N$, $s=0$ would produce an eigenvalue $|\lambda_i | = 1$,
in which case a different small integer must be used.

The straightforward evaluation of the matrix-vector product in equation 
(\ref{eq:eq1}) requires $O(N^2)$ operations to produce $N$ random
numbers, making it hopelessly slow compared with other popular RNG's,
so a faster algorithm would be needed.  
After considerable effort,
they reduced the time to $O(N\ln N)$, much faster but still too slow.

George Savvidy's son Konstantin, also a theoretical physicist,
found the algorithm \cite{sav2014} that reduced
the computation time to $O(N)$, making it competitive in speed
with the fastest generators and, very important, the time to generate 
one number would now be essentially independent of $N$.

Considerable work remained, the most difficult being the determination
of the period of the new generator for all interesting values of N. 
The periods are so long that this requires the use of finite 
(but very large) Galois fields.

By 2014, the essential problems were solved, MIXMAX was being tested
at different sites including CERN, and Konstantin Savvidy published the
paper  \cite{sav2014} 
giving some of the important properties of MIXMAX for 
13 different values of the dimension $N$ from 10 to 3150.   
The implementation uses 64-bit integer arithmetic internally,
so that the user always gets standard double-precision floating-point 
numbers with 53-bit mantissas.  
All the periods are longer than $10^{165}$, so that is not a problem.
The paper gives the results of the Big Crush test of the
        TestU01 package  \cite{testu01}   for 13 values of 
$N$, and the test is failed for $N \leq 64$ 
     \footnote {for $N=64$, the failure of the Big Crush test 
      is termed ``only marginal", but the probability of 
      $\chi ^2  \geq 372$ for 232 degrees of freedom is 
       $1.12 \times 10^{-8}$. }
but passes for $N \geq 88$.  
In practice, $N = 256$ became the default value, probably because of
the results of these tests.

As expected, most of the eigenvalues of the Savvidy matrix are far from
the unit circle (compared with those of RANLUX), 
although there are always a few close to one.  
 For some values of $N$ (including $N=256$)
  the ``magic" integer $s$ must be invoked as described above.
The moduli of the largest eigenvalues are typically of order $N$.
For large $N$, the smallest eigenvalues cluster around $|\lambda| = 1/4$.

Thus the distribution of eigenvalues is more favorable than that of RANLUX, and indeed the Kolmogorov entropy is greater, 
indicating a faster convergence toward the asymptotic independence
that is guaranteed by the mixing property.
In order to demonstrate clearly that mixing is indeed occurring, 
one can plot the divergence of nearby trajectories 
as was done for RANLUX, and verify that the divergence is exponential
as it should be for a continuous K-system.  
Then the same plot can be used to determine how much (if any) 
decimation is required to eliminate the correlation of nearby trajectories.

\begin{figure}

\includegraphics[width=0.6\linewidth]{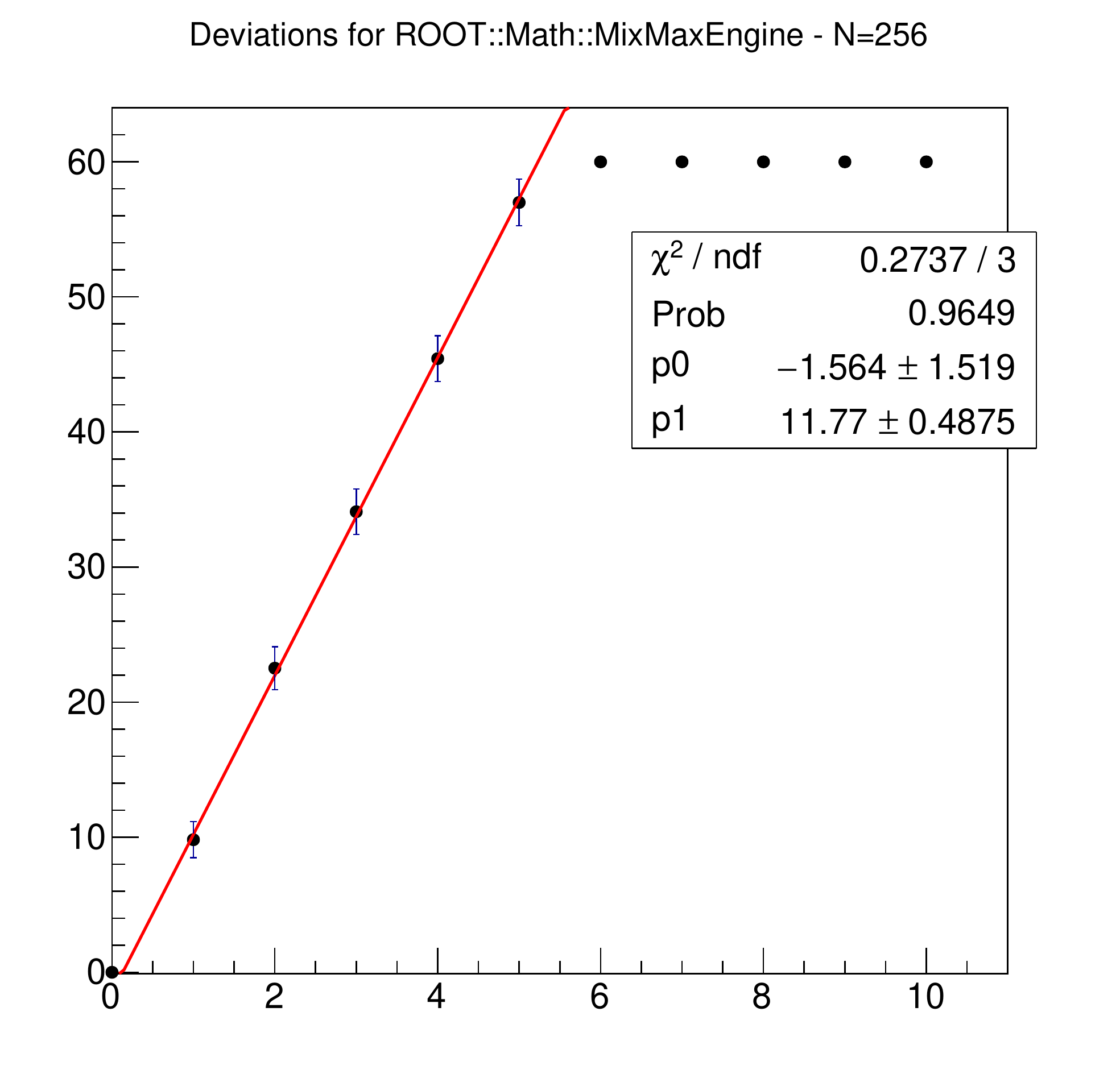}

\caption{
The divergence of nearby trajectories for MIXMAX 1.0, N=256. 
 The y-axis is the distance $\log_2\delta$. }
\label{fig:2}
\end{figure}

Unfortunately, there was nothing in the published paper   \cite{sav2014} 
on the divergence of nearby trajectories and no mention of decimation. 
At CERN, we were mainly interested in getting a RNG 
of the highest quality with no known defects, 
so we did our own calculation
of the divergence of nearby trajectories for $N$=256 
as shown in figure \ref{fig:2}.  
It can be seen that
(1) the trajectories do indeed diverge exponentially, and
(2) after five iterations, the average distance between two nearby
trajectories has almost reached the asymptotic value 
(slightly less than the maximum value of $1/2$).
Note that as long as the average distance between trajectories is
significantly less than the asymptotic value, this is direct evidence of 
a defect, since it indicates
that the trajectories still ``remember where they came from'', 
which would not be the case if the points were independent.
But this generator passed the Big Crush test, so we see clearly
that the divergence of trajectories is more sensitive to this defect than
Big Crush.

So MIXMAX 256 requires decimation, but less than RANLUX,
making it a candidate for the world's fastest high-quality RNG.  
This version was introduced at CERN as a standard generator for 
ROOT, with the possibility to choose different levels of decimation. 

\begin{figure}

\includegraphics[width=\linewidth]{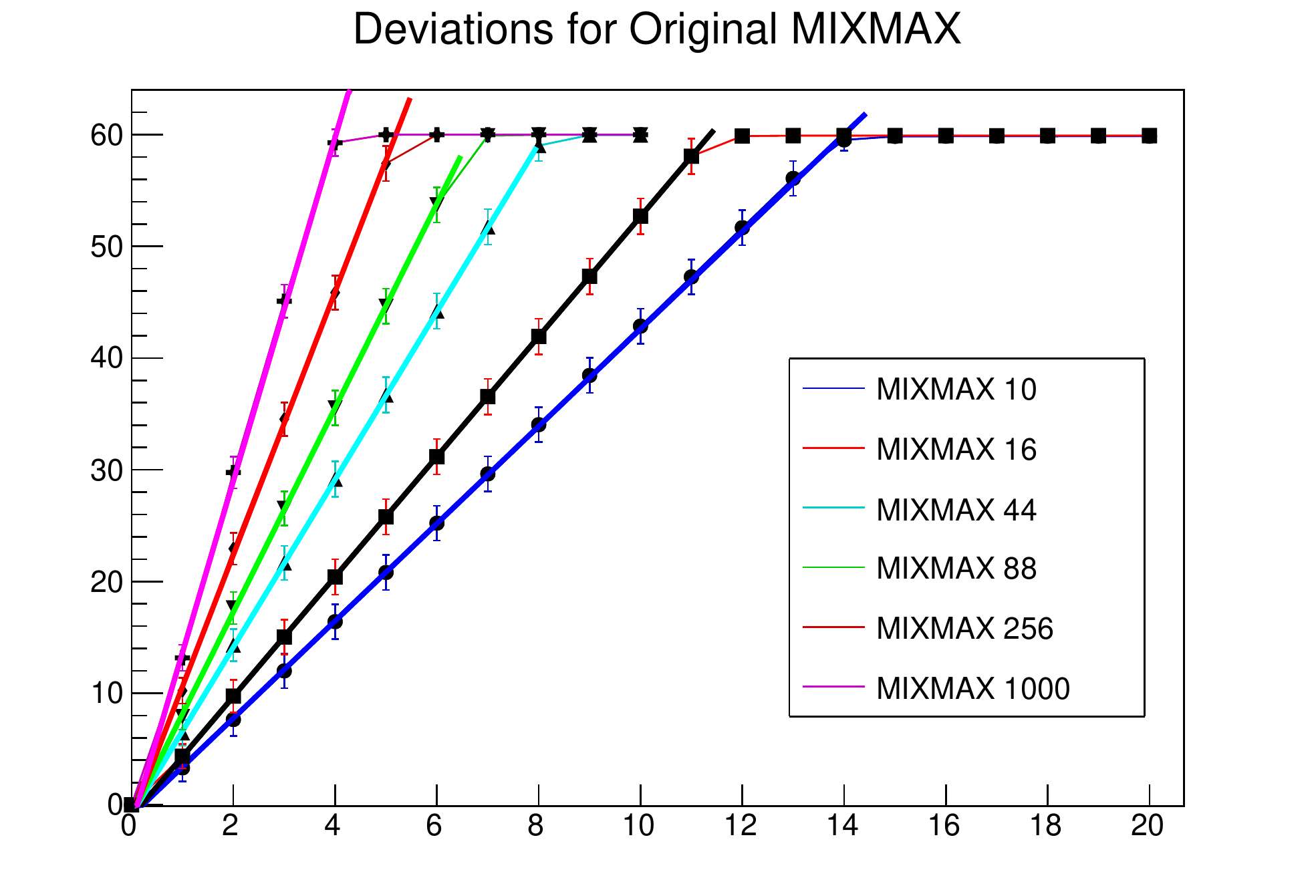}

\caption{
The divergence of nearby trajectories for MIXMAX 1.0, N=
10, 16, 44 ,88, 256, 1000. 
 The y-axis is the distance $\log_2\delta$.}
\label{fig:divall}
\end{figure}

We have also considered the properties of MIXMAX for other values
of the dimension $N$.   In particular, we have looked at the divergence
of nearby trajectories for some values of $N$ for which 
the period was published in  \cite{sav2014} .
These curves are shown in figure \ref{fig:divall} for 
N =10, 16, 44, 88, 256 and 1000.  
(We have also tried N=44000, but the overheads 
involved in handling such a large matrix are not worth the possible
gains.)
It is seen that for all these values of N, the divergence is remarkably
exponential, not only for small separations as required by the theory,
but for all separations right up to the expected asymptotic value
which is always close to $1/2$.
In addition, the slope of the exponential is in all cases equal to
the maximum Lyapunov exponent as predicted by the theory.
And since the Lyapunov exponent increases with $N$,
the decimation required for maximum separation decreases with $N$.
However, there does not seem to be much to be gained in going to
values of $N>256$, especially since the overhead of initialization
and memory usage would then start to become significant.  

    With the possible exception of $N=10$ 
(which some may consider too small)
all the generators shown in figure  \ref{fig:divall} satisfy all our
criteria for highest quality RNG's with no known defects, provided 
of course that appropriate decimation is applied.  This decimation
can be read off the figure and is given in Table \ref{tab:MXdecimation}.

  MIXMAX also offers the possibility to seed the generator at different
faraway points in such a way as to avoid overlapping with sequences 
used in other calculations.  So it satisfies all our criteria for highest
quality as long as appropriate decimation is applied. 

\begin{table}[t]
\centering
\begin{tabular}{|c|c|}
\hline
MIXMAX N  &  decimation \\ \hline
10 & 14  \\
16 & 11  \\
44 &  8  \\
88 &  6  \\
256 &  5  \\
1000  &  4  \\
\hline
\end{tabular}
\caption{Decimation required for full separation 
with original MIXMAX for different values of $N$.}
\label{tab:MXdecimation}
\end{table}

However, the developers of MIXMAX were already working on an extended
MIXMAX which hopefully would \emph{not require any decimation.}

\section{The High-quality RNG's: 3. Extended MIXMAX}
  
The search for a generating matrix with larger entropy (for faster 
convergence to asymptotic mixing)
consists in looking for transformations of $A$ which do not change the 
determinant, and do not interfere with the algorithm for fast multiplication
but do move the eigenvalues further away from the unit circle.
For the purposes of this section, it will be convenient to rewrite
equation (\ref{eq:eq1}) as it is implemented in the computer algorithm,
using explicitly integer arithmetic:
            \begin{equation}  \label{eq:eq1int} 
       a(i+1) =  A \times a(i)  \mod p 
            \end{equation}
where $a(i) $ is now an $N$-vector of 60-bit integers and  
$p=(2^{61}-1) $.  As before, $A$ is an $N \times N$ matrix of integers,
but in extended MIXMAX, some of the integers may be very large.

The simplest modification which increases the entropy is to use much 
larger values of  the magic integer $s$ in (\ref{eq:savmat}) up to 
the largest integers allowed by the computer arithmetic.
This is called in \cite{sav2016}
the two-parameter family $A(N,s)$ with large $s$.

A further modification was introduced using a new integer $m$ to 
produce a new matrix $A$ which is called the three-parameter family
$A(N,s,m)$ (see eq. \ref{eq:savmatnew} ),
   \begin{equation}  \label{eq:savmatnew} 
  A(N,s,m) = 
 \begin{pmatrix}
  1 & 1     & 1 & 1 & \cdots & 1 & 1 \\
  1 & 2     & 1 & 1 & \cdots & 1 & 1 \\
  1 & m+2+s & 2 & 1 & \cdots & 1 & 1 \\
  1 & 2m+2     & m+2 & 2 & \cdots & 1 & 1 \\
  1 & 3m+2     & 2m+2 & m+2 & \cdots & 1 & 1 \\
  \vdots  & \vdots  & \vdots & \vdots & \ddots  &\vdots &  \vdots &\\
  1 & (N-2)m+2 & (N-3)m+2 & (N-4)m+2 & \cdots & m+2 & 2 \\
 \end{pmatrix}   
  \end{equation}
where it can be seen that for $m=1$, $ A(N,s,m)$ reduces to 
the two-parameter $A$.

The main properties of the new generator were published 
in \cite{sav2016} for selected values of $(N,s,m)$ for which 
the period could be determined and which would be of most interest.  
Some of the results were nothing short of spectacular.

Probably the most impressive was the spectrum
of eigenvalues for the parameter values: \mbox{$N$=240,} 
\mbox{m=$2^{51}$+1,}
and s= 487013230256099140, for which all the eigenvalues 
have modulus
around $10^{15}$ except for one which is so close to zero that the
product of all of them is one, as it must be.

\begin{figure}  
\includegraphics[width=\linewidth]{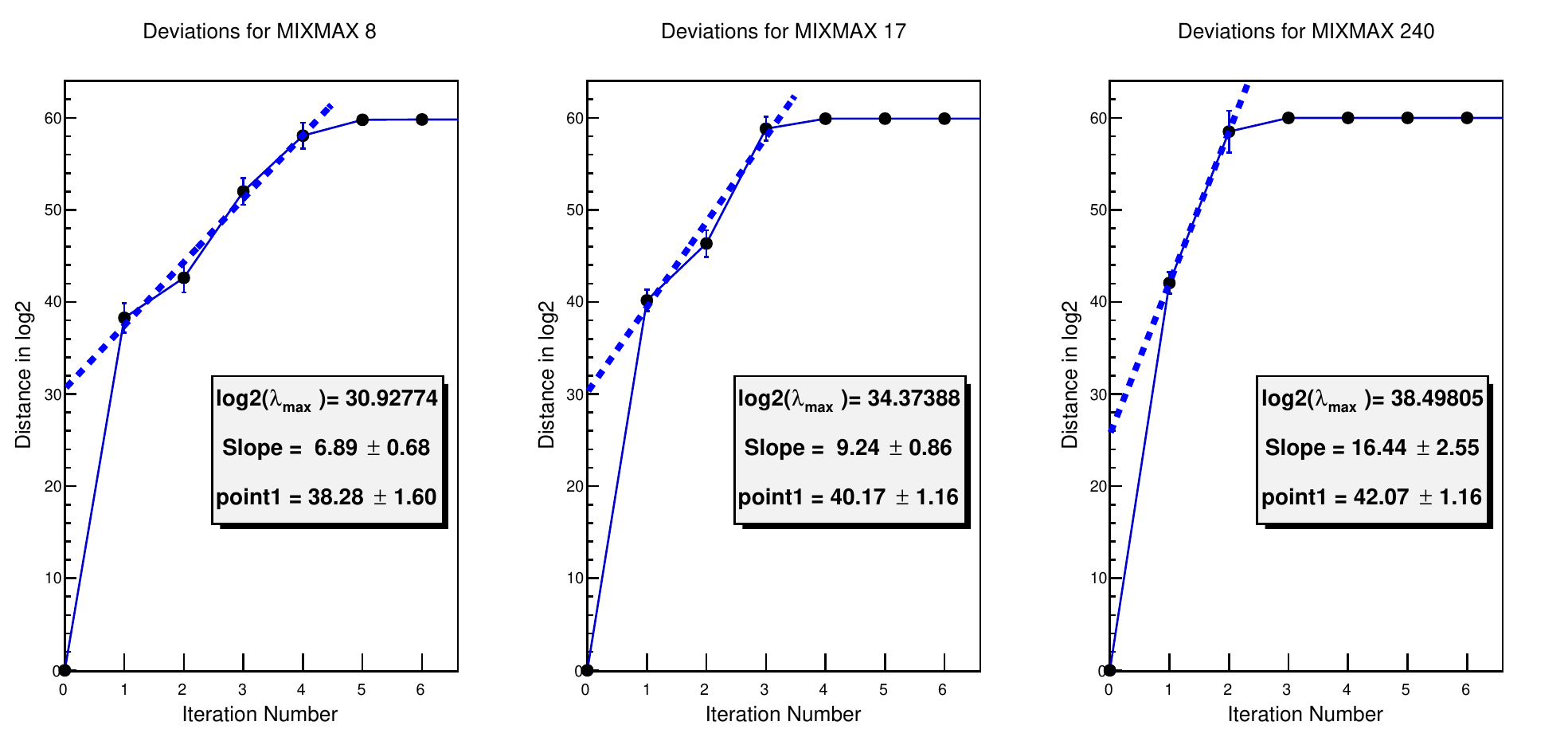}
\caption{The divergence of nearby trajectories for extended MIXMAX
with parameters recommended in \cite{martir}.
 On the left, for $N=8, m\approx 2^{36}$.
 In the middle, for $N=17, m\approx 2^{36}$.
 On the right,  for $N=240, m\approx 2^{32}$. }
\label{fig:3}
\end{figure}

However, as was the case for the original MIXMAX, the authors did not
publish any results concerning the divergence of nearby trajectories
for these generators.   So we calculated this ourselves 
for some values we considered implementing.  
The results are shown in figure  \ref{fig:3} for three interesting values
of the parameters $(N,s,m)$ which are recommended 
by the MIXMAX developers in \cite{martir}.

 It is clear from figure \ref{fig:3} that none of the extended MIXMAX
 generators we considered exhibit exponential divergence 
 as they should for a K-system and as they do for original MIXMAX.  
 Furthermore, none of the ``slopes" that one could identify
 on these curves corresponds to the known Lyapunov exponent
 as was the case for the original MIXMAX.
 We have also made the same plots for extended MIXMAX for  even
 larger values of $m$ (but still keeping  $Nm\!< \! p$),
 and these plots show a divergence so fast that it reaches the
 asymptotic value on the first step, so it is impossible to see whether
 the divergence is exponential.
 
     We have tried to find the reason for this behaviour.
The obvious suspect is the large values of some elements of $A$
 which have caused the extended algorithm to lose the mixing 
 properties of the continuous K-system.  
 One upper limit on the allowed magnitude of elements of $A$
 is given by the developers of MIXMAX in \cite{sav2016}:

 \begin{quote}
 It is most advantageous to take large values of $m$, but preferably 
 keeping $Nm\!< \! p$, such as to have an unambiguous correspondence
 between the continuous system and the discrete system 
on the rational sublattice.
 \end{quote}
 The condition $Nm\!< \! p$ is certainly necessary as indicated 
 in the above quotation,
 but the values of $mN$ in our examples already satisfy that
 criterion, so there must be an additional explanation having to do 
 with the multiplication of large integers (modulo p), 
since the problem does not arise when the $A_{ij}$ are small.
Consider for example just one 
component of the matrix-vector multiplication, that is, how the $i$th 
component of the new vector $a^+$ depends on the $j$th 
component of the previous vector $a$:
    $$a_i^+  = A_{ij}  a_j  \mod p $$
\noindent
When the integer $A_{ij}$ is very big, this equation describes a
trajectory that loops rapidly around the torus, 
and when $A_{ij}$ is equal to $p$, 
there is only one discrete point on each loop of the continuous 
trajectory around the torus.  Clearly one point is not 
even enough to ``see" whether or not the continuous
trajectory goes around the torus, but how many points 
on each turn would be required in order to represent accurately
the continuous trajectory?

For the complete trajectory in 240 dimensions it becomes 
difficult to determine what values of $A_{ij}$ are small enough.  
Fortunately we can again make use of the separation
of nearby trajectories, which must be exponential for the continuous 
system, so it should also be exponential for a discrete system
which is a good approximation to the continuous system.

The lack of exponential divergence in figure  \ref{fig:3}  means that
extended MIXMAX with the given values of $N,m,s$ does not
meet our criteria for K-mixing.

\subsection{Extended MIXMAX: The Spectral Test and Skipping.}    
Although originally the spectral test was applicable only to simple LCG's,
it has been extended more recently to more complex LCG's, and
L'Ecuyer, Wambergue and Bourceret have applied it 
\cite{MXspectral}
to extended MIXMAX
which can be analyzed as a matrix LCG.
They find that the lattice structure in certain dimensions 
(notably in 5 dimensions for $N=240$ and $m=2^{51}+1 $) 
is ``very bad", all points lying on a number of hyperplanes much
smaller than should be the case for their $p=2^{61} - 1$.
They also point out the existence of ``simple linear relations" 
between coordinates of successive points which indicate unwanted 
correlations.   
Both problems remain even when the guilty coordinates  are
skipped in the output.

    These problems were known to us but we did not consider them 
serious because they were both eliminated (at least in RANLUX) 
by decimation as described above in this paper.  Note that
the skipping of coordinates performed by L'Ecuyer et. al. 
in \cite{MXspectral}  (and also by the MIXMAX developers 
in \cite{martir}) is very 
different from the decimation proposed in RANLUX.
L'Ecuyer et. al. discard certain coordinates of every vector
generated, whereas L\"uscher discards whole vectors at a time.
The latter is justified by the asymptotic nature of K-mixing so the fact
that it is effective for RANLUX is further indication
that K-mixing is indeed occurring.
For extended MIXMAX, we cannot use the theory of
K-mixing to justify decimation.

\section{The High-Quality RNG's: 4. RANLUX++}
As mentioned above in connection with the spectral test,
the RANLUX algorithm is known to be
equivalent to a linear congruential generator (LCG) with an enormous
multiplier.   
This fact was used by L\"uscher \cite{luscher} in order to apply the 
spectral test to RANLUX, 
but otherwise seemed to be of little practical interest since
the computers of that time were unable to do long multiplications 
fast enough to be useful.  More recently however, Sibidanov 
\cite{sibidanov}  has 
produced a version called RANLUX++, implemented using the linear
congruential algorithm with a clever way of doing long integer 
multiplication on modern computers, which makes it faster than
standard RANLUX and does the decimation at the same time. 
The algorithm of Sibidanov cannot be programmed efficiently 
in the usual high-level languages, but is possible in 
assembler code using operations he says are now available on 
most computers used for scientific computing.

The heart of the new software is the procedure to multiply two 576-bit 
integers using 81 multiplications of 64 x 64 bits cleverly organised to be
extremely fast using modern extensions of arithmetic and fetching 
operations described in Sibidanov \cite{sibidanov}.  
For our purposes, there are two important features of this method:
First, that it is able to reproduce the sequences of RANLUX 
much faster than the high-level language versions, and second, 
that it can produce more decimation (higher luxury levels) than the 
predefined levels in RANLUX, without increasing the generation time.

In order to understand the importance of this second feature, we should
look more closely at the mechanism used to implement RANLUX
with the LCG algorithm.  
Using the notation of Sibidanov, one random number $ x_{i+1}$ is generated by
            \begin{equation}  \label{eq:lcg} 
       x(i+1) =  a \cdot x(i) + c \mod m 
            \end{equation}
With the parameters corresponding to RANLUX, the base $b = 2^{24}$, 
 the modulus $m=b^{24} - b^{10} + 1$, 
 the multiplier $a = m - (m-1)/b$, and $c=0$.
 
    Decimation at level $p$ in RANLUX is implemented by
delivering 24 numbers to the user, 
then throwing away $p- 24$ numbers before delivering the next 24.
In RANLUX++, the decimation is called ``skipping"  \cite{sibidanov}
because it is implemented directly by 
            \begin{equation}  
       x(i+p) =  \xapm \cdot x(i)  \mod m ,
            \end{equation}
where $\xapm$ is a precomputed very long integer constant.
As expected, this makes generation in RANLUX++ much faster than 
that of the original RANLUX.  
But the surprise comes when we look at the values of \mbox{$\xapm$}
for different values of $p$ corresponding to the standard luxury levels,
as well as two new higher levels, $p=1024$ and $p=2048$,
shown in Table 1 of  \cite{sibidanov}.

Here we see that as the luxury level increases, the obvious
regularity in the factor $\xapm$ decreases, but is still clearly present
for the value of $p$ corresponding to the highest 
standard luxury level of RANLUX, $p=389$.  
\footnote{
The values of $p$ corresponding to different luxury levels have changed,
so the value used by Sibidanov (389) is slightly different from the 
value for the current highest luxury level for the 24-bit algorithm (397), 
but the difference is not significant.}
Sibidanov remarks that
apparently, it would be advantageous to apply still more decimation,
at least to $p=1024$ or $p=2048$, which can be obtained ``for free".
In this context, we have applied a simple runs test to the hex patterns
for the factors $\xapm$ for the two new values of $p$ and find that they 
both pass the test for expected numbers of 2-runs, 3-runs and higher,
which would indicate that the patterns are already as random for 
$p=1024$ as expected for a truly random sequence of 144 hex digits.
So, if the randomness of $\xapm$ is an indication of the degree of mixing,
$p=389$ is not enough, but $p=1024$ is already fully mixing.

\section{Conclusions}
We now have several high-quality RNG's available for general use
based on the theory of Mixing in classical dynamical systems.

\subsection{RANLUX}

This easily portable and well-documented RNG exists in both 
single- and double-precision versions both of which satisfy 
our criteria for K-mixing.  
At the highest standard luxury level 
(currently level 2, p=397 for the single-precision version, 
p=794 for the double-precision version),
 it has for many years been
considered the most reliable RNG available, but too slow for
some applications.  Recent improvements in code optimization 
have speeded it up considerably, so it should now be acceptable
for most applications (see timings below).

However, new evidence from RANLUX++ indicates that 
even more decimation up to p=1024 
may be needed to eliminate the higher-order correlations. 
This level of decimation would make RANLUX slower,
but is obtained ``for free" with RANLUX++.

\subsection{MIXMAX }

We consider only  the original MIXMAX with small $s$ and no $m$, 
described in \cite{sav2014}.
This RNG produces always double-precision floating-point numbers,
and the user has the freedom to choose $N$,
the size of the generating matrix $A$.  In practice one will choose
a  value of $N$ for which the period is known,
tabulated in \cite{sav2014}.
The code is maintained and available under HepForge.

For users who require the highest quality random numbers, 
it will be necessary to introduce into MIXMAX the possibility
to decimate the appropriate number of iterations for each vector 
of random numbers delivered to the user, which is not difficult 
but requires modifying the source code.  
This has already been done for the version with N=256 installed 
at CERN.  
The decimation required for other values of $N$ is given here in 
Table \ref{tab:MXdecimation}.

Assuming the user has chosen a value of $N$ and has
introduced appropriate decimation according to the criteria given here
for RANLUX, he or she is still faced with the problem whether
additional decimation may be useful to eliminate higher order 
correlations which have never been observed.
But this ``problem" could actually turn out to be an advantage in case 
of doubt whether the RNG is good enough for the application at hand:
the calculation can be repeated with even higher degree of decimation
and if the result is significantly different, we can use the higher degree.
The theory
guarantees that the higher decimation should only improve the
mixing if it is not yet complete.

\subsection{RANLUX++ }
This reincarnation of RANLUX by Alexei Sibidanov looks like it is going 
to be hard to beat, since it is the LCG equivalent of the RANLUX 
algorithm with even more decimation and runs faster than any other
RNG considered here. 
Although we have as yet little experience using it, 
the only drawback we can see is that parts of it cannot be coded 
in the usual high-level languages because it requires access to
arithmetic instructions available only in assembler code.
But the author claims these instructions are available on all the
computers commonly used for MC calculations, so that the
individual physicist should be able to obtain a copy that will
run on his/her machine.  

\subsection{Timing Benchmarks}
We think that the overriding consideration in choosing a RNG should 
be the quality of  the random numbers in order to reduce to a minimum 
the risk of obtaining an incorrect result  which may never be
recognised as such.
However, since some applications may require in addition a high speed
of generation, we give some typical timings we have observed for
the RNG's described here which meet our criteria for highest quality
and no known defects.  
That means for RANLUX the highest standard luxury level and for
MIXMAX 256 decimation ``skip 5".  
For RANLUX++ the decimation is even much higher since it is 
obtained ``for free" as described above.

The table below gives our timings in nanoseconds per
random floating-point number between zero and one 
observed on three different platforms A, B, C.

\bigskip
\noindent
\begin{tabular}{l l c c c}
\hline
Single Prec. &   \ & A & B & C  \\  
       \                   &RANLUX S   &  17.6  &  18.8  & 36.4   \\
       \                   &RANLUX++  &  n.a.   &   5.9   &  10.2  \\  \hline
 Double Prec.     & \ & \ & \ &        \\  
       \                   &MIXMAX 256 & 17.0 & 18.2 & 31.3   \\
       \                   &RANLUX D    & 31.4 & 34.4 & 67.5   \\
       \                   &RANLUX++   & n.a.  & 7.9    & 13.7  \\
\hline  \\
\end{tabular}
Platform A is a MacBook i7, 2.6GHz;
 B is a desktop i9-9900K, 3.7GHz; and C is a server Xeon E5-2683, 
 2 GHz.  RANLUX++ did not run on our MacBook.
RANLUX is often used with SSE extensions which are foreseen 
in the standard code and speed it up by about 30\% but are not
portable.

\section*{Acknowledgements}
We have benefitted greatly from discussions over many years
with Martin L\"uscher and George Savvidy, and we are also grateful
for contributions from Konstantin Savvidy, John Harvey, John Apostolakis,
and Philippe De Forcrand.

\end{document}